\documentclass[12pt]{article}
\usepackage[margin=1in]{geometry}

\usepackage{graphicx,verbatim,array,multicol, subfigure, xcolor, lscape, mathrsfs}
\usepackage{psfrag, amsmath, amsfonts, epsfig, fancybox, setspace,soul,amsthm,natbib,pdflscape}
\usepackage[shortlabels]{enumitem}
\usepackage[hidelinks]{hyperref}
\def\supg{^{(g)}}

\def\bX{{\bf X}}

\def\bw{{\bf w}}
\def\bx{{\bf x}}

\def\supone{^{(1)}}
\def\supzero{^{(0)}}
\def\supg{^{(g)}}

\def\var{\text{Var}}

\begin{document}

\begin{center}
\LARGE{Assessing Surrogate Heterogeneity in Real World Data Using Meta-Learners}\\

\vspace*{15mm}
\normalsize Rebecca Knowlton$^{1}$ and Layla Parast$^{1}$ \\
\vspace*{10mm}
\normalsize $^{1}$Department of Statistics and Data Sciences, University of Texas at Austin\\
\end{center}

\clearpage
\begin{abstract}
Surrogate markers are most commonly studied within the context of randomized clinical trials. However, the need for alternative outcomes extends beyond these settings and may be more pronounced in real-world public health and social science research, where randomized trials are often impractical.  Research on identifying surrogates in real-world non-randomized data is scarce, as available statistical approaches for evaluating surrogate markers tend to rely on the assumption that treatment is randomized. While the few methods that allow for non-randomized treatment/exposure appropriately handle confounding individual characteristics, they do not offer a way to examine surrogate heterogeneity with respect to patient characteristics.  In this paper, we propose a framework to assess surrogate heterogeneity in real-world, i.e., non-randomized, data and implement this framework using various meta-learners. Our approach allows us to quantify  heterogeneity in surrogate strength with respect to patient characteristics while accommodating confounders through the use of flexible, off-the-shelf machine learning methods. In addition, we use our framework to identify individuals for whom the surrogate is a valid replacement of the primary outcome. We examine the performance of our methods via a simulation study and application to examine heterogeneity in the surrogacy of hemoglobin A1c as a surrogate for fasting plasma glucose. 
\end{abstract}


\clearpage
\section{Introduction}
The increased use of surrogate markers has been an important advancement in clinical trials, offering a pathway to more efficient and cost-effective research for complex diseases like cancer and AIDS \citep{katz2004, fleming1994}. A surrogate marker is formally defined as a person-level measure that serves as a substitute for a direct measure of a primary outcome, facilitating the evaluation of treatment or exposure effects. While surrogate markers are most commonly studied within the context of randomized clinical trials, the need for alternative outcomes extends beyond these settings. In fact, this need may be even more pronounced in non-randomized studies. In real-world public health and social science research, where randomized trials are often impractical or unethical, surrogate markers may play a crucial role in enabling timely decision-making about treatment or exposure effects \citep{rosenbaum2005observational, boyko2013observational}.

\subsection{Related Work}
Research on identifying surrogates in real-world data (i.e., not randomized) is scarce, as available statistical approaches for evaluating surrogate markers tend to rely on the assumption that treatment is randomized. Recently, \citet{han2022identifying} proposed an approach to identify surrogate markers in real-world data by quantifying surrogate strength using the proportion of the treatment effect (PTE) on the primary outcome that is explained by the treatment effect on the surrogate with estimation using inverse probability weighting and doubly robust estimators. \citet{agniel2023doubly} offered a flexible doubly robust method to estimate the PTE of a high-dimensional surrogate in a non-randomized setting with implementation via the relaxed lasso and the super learner \citep{meinshausen2007relaxed,van2007super}. \cite{agniel2024robust} recently extended this approach to a longitudinal surrogate with a censored time-to-event outcome through the use of efficient influence functions for the treatment effect estimands, with implementation using a one-step plug-in estimator and a targeted minimum loss-based estimator \citep{rose2011introduction}. 

These methods are useful for settings where treatment is not randomized and thus, one must account for individual characteristics which may be potential confounders. However, these methods do not offer a way to examine surrogate heterogeneity with respect to patient characteristics. Similar to (but different from) the idea of treatment effect heterogeneity, surrogate heterogeneity means that the surrogate may be useful, i.e., a valid replacement for the primary outcome, for some individuals but not others \citep{parast2021hetero}. Of course, this is especially problematic if the surrogate is then used to replace the primary outcome in a future study, which is the ultimate goal of surrogate identification. Specifically, one may end up using a surrogate to make a decision about the effect of a treatment or exposure in a future study when in fact, the surrogate is a poor replacement of the primary outcome for the individuals in that study \citep{parast2022using}. Recent work has offered methods to assess and test for surrogate heterogeneity, but they have been limited to randomized settings. For example, \citet{roberts2021incorporating} offered a Bayesian-based approach for surrogate validation conditional on baseline covariates in a principal stratification framework within a randomized setting. Also within a randomized setting, \citet{knowlton2024} and \citet{parast2021hetero} proposed flexible approaches to estimate the PTE of a surrogate as a function of baseline covariates and formally test for evidence of heterogeneity. 

To our knowledge, there do not exist any methods to assess heterogeneity in the PTE of a surrogate in a non-randomized setting. In this paper, we aim to fill this gap by proposing a framework to assess surrogate heterogeneity in real-world (non-randomized) data and implement this framework using various meta-learners. Our approach allows us to quantify surrogate strength and assess potential heterogeneity in surrogate strength with respect to patient characteristics while accommodating confounders through the use of flexible, off-the-shelf machine learning methods. In addition, we use our framework to identify individuals for whom the surrogate is a valid replacement of the primary outcome, that is, individuals for whom the proportion of the treatment effect explained by the surrogate is greater than some prespecified threshold.

\subsection{Organization of the Paper}
The paper is organized as follows. In Section \ref{setting} we describe our notation, setting, assumptions, and proposed framework. In Section \ref{estimation} we propose various T-learner estimation methods including simple linear estimation, generalized additive model (GAM) estimation, and estimation via regression forests. In Section \ref{identify} we propose a procedure to use our resulting estimates to identify individuals for whom the surrogate is sufficiently strong. We examine the performance of our proposed methods using a simulation study in Section \ref{sims} and apply the methods to examine heterogeneity in the surrogacy of hemoglobin A1c as a surrogate for fasting plasma glucose in an observational data set in Section \ref{example}. 

\section{Setting and Proposed Framework}\label{setting}

\subsection{Notation, Setting, and Assumptions} 

Let $G$ denote the treatment or exposure, where $G=1$ indicates the treatment group and $G=0$ indicates the control group, or a comparative treatment. Since the real-world data are observational, treatment is \textit{not} randomly assigned at baseline. Let $\mathbf{X}$ denote a $p$ dimensional vector of baseline variables, $S$ denote the surrogate marker measured after baseline, and $Y$ denote the primary outcome of interest measured after baseline.  Under the potential outcomes framework, we consider $S^{(g)}$ and $Y^{(g)}$, which denote the surrogate marker and primary outcome values under treatment $G=g$, respectively. The full potential data set thus encompasses $(Y^{(1)}, Y^{(0)}, S^{(1)}, S^{(0)}, \bX)$, though we observe either $(Y^{(1)}, S^{(1)}, \bX)$ or $(Y^{(0)}, S^{(0)}, \bX)$ for each subject, contingent on the treatment received. Therefore, the observed data consists of independent and identically distributed (iid) copies of $(Y^{(1)}, S^{(1)}, \bX)$ for the treatment group, denoted $(Y_{1i}, S_{1i}, \bX_{1i})$ for $i=1,\ldots,n_1$, and iid copies of $(Y^{(0)}, S^{(0)}, \bX)$ for the control group, denoted $(Y_{0i}, S_{0i}, \bX_{0i})$ for $i=1,\ldots,n_0$. Here, $n_g$ represents the number of individuals in treatment group $g$, and the total sample size is $n=n_0+n_1$.

We first require a number of strong but common untestable causal assumptions:

\begin{enumerate} 
\item[] (C1) [Consistency] $Y\supg = Y$ and $S\supg = S$ when $G = g$;
\item[] (C2) [Positivity/Overlap]
$P\{\pi_g(\bX) > \epsilon\} = 1$, where $\pi_g(\bx) = P(G = g | \bX = \bx)$, for some $\epsilon > 0$, and $f(S\supg | \bX=\bx)>0$ for $g = 0,1$;
\item[] (C3)[Unconfoundedness] $Y\supg, S\supg \perp G | \bX~$ and $~Y\supg \perp S\supg | G, \bX$;
\item[] (C4) $Y\supone \perp S \supzero |S \supone, \bX$ and $Y\supzero \perp S \supone |S \supzero , \bX$; and
\item[] (C5) $E(Y\supg| \bX= \bx)$ and $E(Y\supg| S\supg, \bX= \bx)$ for $g = 0,1$ are Lipschitz continuous.
\end{enumerate} \vspace{-2mm}

\noindent Assumption (C1) states that the observed outcome and surrogate under treatment $g$ are equal to their potential outcomes when treatment $G=g$ is actually received. Assumption (C2) states that for any $\bx$, there is a positive probability of receiving each treatment and ensures overlap in the support of $S\supone$ and $S\supzero$. Assumption (C3), referred to as unconfoundedness, first states that treatment assignment is independent of potential outcomes and potential surrogate values, conditional on observed covariates. In particular, this requires no unmeasured confounding between treatment and either the surrogate or the outcome. The second component of Assumption (C3) requires no unmeasured confounding of $(Y\supg, S\supg)$ given the observed covariates and treatment group $G$. Assumption (C4), similar to the assumption of sequential ignorability and cross-world independence \citep{imai2010general,andrews2021insights},  states that given the surrogate under one treatment assignment and the covariates, the outcome under that treatment is independent of the surrogate value under the other treatment.  Lastly, Assumption (C5) is needed to ensure certain asymptotic properties, discussed in Section \ref{inference}. While (C2) may be explored to some extent empirically, the other assumptions rely on potential outcomes that are not testable from observed data alone. Assumption (C4) in particular is a strong assumption that is difficult to verify, as it involves potential outcomes under different treatments that are never simultaneously observed.

\subsection{Proportion of Treatment Effect Explained} 

In this paper, the measure of surrogate strength that we focus on is the proportion of the treatment effect on the primary outcome that is explained by the treatment effect on the surrogate marker, which if often abbreviated as PTE \citep{wang2002,freedman1992}. The PTE is a single number summary defined based on contrasts between the overall treatment effect and the residual treatment effect after accounting for the effect on the surrogate. Here, we first describe this quantity as proposed in \citet{wang2002}, which ignores potential heterogeneity and assumes randomized treatment. In the following section, we will build from this definition specifically incorporating heterogeneity and removing the randomization assumption. The overall treatment on $Y$ is defined as $$\Delta = E(Y \supone - Y \supzero)$$ and the residual treatment effect is defined as
\begin{eqnarray*} 
\Delta_S &=& \int E(Y \supone - Y \supzero \mid  S \supone = S \supzero = s) dF_{S^{(0)}}(s), \end{eqnarray*}  
where $F_{S^{(0)}}(\cdot)$ is the marginal cumulative distribution function of $S \supzero$. The residual treatment effect conceptually measures the treatment effect on the primary outcome that remains after adjusting for the treatment effect on the surrogate. Using these quantities, the proportion of the treatment effect explained is defined as $R_S = (\Delta-\Delta_S)/\Delta = 1 - \Delta_S / \Delta$. In general, high values of $R_S$ indicate that a high proportion of the treatment effect is explained by $S$ and thus, S is a strong surrogate, while lower values of $R_S$ indicate a poor surrogate; we expand on this in Section \ref{identify}.

\subsection{Proposed Framework to Assess Surrogate Heterogeneity} \label{proposed}

Building from the standard PTE definition, we now define: \begin{eqnarray*}
\Delta(\bx) &=&  E(Y \supone \mid \bX=\bx) - E(Y \supzero\mid \bX=\bx), \quad \mbox {and} \\
\Delta_S(\bx) 
&=& \int E(Y \supone - Y \supzero \mid  S \supone = S \supzero = s, \bX=\bx) dF_{S^{(0)}|\bX}(s), 
\end{eqnarray*}  
where $F_{S^{(0)|\bX}}(\cdot)$ is the conditional CDF of $S \supzero|\bX = x$, and we define the PTE as a function of $\bX=\bx$, so that $R_S(\bx) = 1 - \Delta_S(\bx) / \Delta(\bx)$. Throughout, we assume that $\Delta(\bx) \neq 0 ~\forall \bx$ to ensure that $R_S(\bx)$ is well-defined.

We first consider $\Delta(\bx)$, which is the conditional average treatment effect (CATE). By Assumptions (C1)-(C3), 
\[
\Delta(\bx) = E(Y \mid \bX=\bx, G=1) - E(Y \mid \bX=\bx, G=0),
\]
which is identifiable from the available data. The problem of CATE estimation is a well-known problem and has received considerable attention in recent literature \citep{athey2019estimating,athey2016recursive,wager2018estimation,caron2022estimating,kunzel2019metalearners}. Classical nonparametric approaches to estimate CATE such as nearest neighbor matching or kernel methods suffer from the curse of dimensionality when the data has more than a couple of covariates, making them impractical for many modern applications. Particularly in our setting, the complexity of the covariates is a significant challenge since they may act both as confounders of treatment assignment and informative of the surrogate strength. Modern approaches that can accommodate higher covariate dimensions and maintain the flexibility of nonparametric approaches include ``Meta-Learners'', for example, S-learners and T-learners \citep{kunzel2019metalearners, caron2022estimating}. A meta-learner is simply the result of combining multiple ``base learners"---which can be any supervised learning or regression estimators---in a specific way to estimate the quantity of interest, while allowing the base learners to take any form. So-called ``S-learners" fit a \ul{S}ingle learner that includes treatment assignment as a predictor, while ``T-learners" instead fit separate learners for each treatment group, i.e., \ul{T}wo distinct learners. We focus here on T-learners due to their flexibility in capturing treatment effect heterogeneity without imposing structural assumptions about how treatment modifies the outcome-covariate relationship.  To implement a T-learner for $\Delta(\bx)$, we require a decision for base learner for the conditional expectation of the outcome given the covariates in each treatment group, which we denote as $$\lambda_g(\bx) = E(Y \supg \mid \bX = \bx) = E(Y \mid \bX=\bx, G=g),$$
where the second equality follows under Assumptions (C1)-(C3). Once the base learner is selected and used to estimate $\lambda_g(\bx)$, we denote the resulting estimates as $\widehat{\lambda}_0(\bx)$ and $\widehat{\lambda}_1(\bx)$, and estimation of CATE is straightforward: $\widehat{\Delta}(\bx) = \widehat{\lambda}_1(\bx) - \widehat{\lambda}_0(\bx).$ 

Next, we consider estimation of the residual treatment effect $\Delta_S(\bw)$, which is not as straightforward. Under Assumptions (C1)-(C4), we have 
\begin{eqnarray*}
    \Delta_S(\bx) &=& \int E(Y  \mid  S  = s, \bX=\bx,G=1) dF_{S|\bX, G=0}(s) \\ 
    && \hspace{0.5in} - \int E(Y \mid  S = s, \bX=\bx, G=0) dF_{S|\bX, G=0}(s) \\ 
    &=& \int \mu_1(s,\bx) dF_{S|\bX, G=0}(s) - \int \mu_0(s,\bx) dF_{S|\bX, G=0}(s),
\end{eqnarray*}
where $\mu_g(s,\bx)= E(Y\supg \mid S\supg = s, \bX = \bx)$ represents the conditional mean function, and where $F_{S|\bX, G=0}(\cdot)$ represents the cumulative distribution function of $S\supzero$ given $\bX=\bx$. Similar to $\Delta(\bx)$, we propose to implement a T-learner for $\Delta_S(\bx)$, but we now require a selection of base learners for both $\mu_g(s,\bx)$ and $\zeta_0(\bx) = E(S\supzero \mid \bX = \bx)$.  

In the following section, we use sample-splitting to obtain these estimates and provide details for the implementation of both T-learners to obtain estimates of $\Delta(\bx)$, $\Delta_S(\bx)$, and $R_S(\bx)$ using three sets of base learners: a linear model, a generalized additive model (GAM), and a regression forest.

\noindent \textit{Remark.} Note that the construction of $R_S(\bx)$ does not inherently impose the constraint that $R_S(\bx) \in [0,1]$, meaning there is no requirement that $0 \leq \Delta_S(\bx) \leq \Delta(\bx)$. This issue has been explored in greater detail in other works, such as \citet{stijven2024proportion} in the surrogate setting, where it is argued that values exceeding 1 can still be meaningful, and more broadly in \citet{preacher2011effect}, which examines methods for decomposing effects into direct and indirect components. In fact, the PTE can extend beyond the $[0,1]$ range unless additional constraints are imposed. One sufficient set of assumptions that ensures $R_S(\bx) \in [0,1]$ aligns with those preventing the surrogate paradox, which is discussed in more detail in Section \ref{discussion}. However, we do not explicitly impose these assumptions here and, therefore, do not require $R_S(\bx)$ to remain strictly within the unit interval.

\section{Implementation and Inference \label{estimation}}

\subsection{Implementation via Metalearners}
Implementation of our framework requires selecting base learners for the following components: the outcome models $\lambda_g(\bx)$, the outcome-surrogate conditional models $\mu_g(s,\bx)$, and the surrogate model in the control group $\zeta_0(\bx)$. Many reasonable  choices exist for estimating these components, from simple, computationally efficient models to more complex, flexible models. We focus on using three sets of base learners---a linear model, a GAM, and a regression forest---via the following algorithm.

\noindent \underline{Algorithm for Estimating $R_S(\bx) $} \\
\noindent \textbf{Step 1: Split the Data.} Split the available data set into a training set that will be used to build the base learners for \(\lambda_g(\bx)\), \(\mu_g(s, \bx)\), and \(\zeta_0(\bx)\), and a testing set that will be used to obtain predictions from the trained learners and estimate \( \Delta(\bx) \), \( \Delta_S(\bx) \), and \( R_S(\bx) \). Use of sample-splitting aims to prevent overfitting and ensure honest assessment of the method's performance on unseen data.\\
\noindent \textbf{Step 2: Select a Base Learner.} Choose a supervised learning method (e.g., linear model, GAM, regression forest). \\
\noindent \textbf{Step 3: Estimate the Conditional Average Treatment Effect, $\Delta(\bx)$.} 
\begin{enumerate}[label=(\alph*)]
    \item \textit{Fit Learners for Each Group (T-Learner)}: Using the selected learner, build a learner using the training set for \(\lambda_g(\bx) = E(Y^{(g)} \mid \bX = \bx)\), for \(g = 0, 1\).
    \item \textit{Predict \(\lambda_g(\bx)\)}: Obtain predictions \(\widehat{\lambda}_0(\bx)\) and \(\widehat{\lambda}_1(\bx)\) from the fitted learners using the testing set.
    \item \textit{Estimate \( \Delta(\bx) \)}: Compute the CATE as: $
        \widehat{\Delta}(\bx) = \widehat{\lambda}_1(\bx) - \widehat{\lambda}_0(\bx).$
\end{enumerate}

\noindent \textbf{Step 4: Estimate the Residual Treatment Effect, $ \Delta_S(\bx)$.}
\begin{enumerate}[label=(\alph*)]
    \item \textit{Fit Learners}: Using the selected learner, build learners using the training set for \(\mu_0(s, \bx)\), \(\mu_1(s, \bx)\), and \(\zeta_0(\bx)\).
    \item \textit{Predict \(\widehat{\zeta}_0(\bx)\)}: For each \(\bx\) in the testing set, predict \(\widehat{\zeta}_0(\bx)\), the expected value of \(S^{(0)}\) given \(\bX = \bx\).
    \item \textit{Evaluate \(\widehat{\mu}_g(s, \bx)\) at \(\widehat{\zeta}_0(\bx)\)}: Using a plug-in estimator, use the testing set to predict $\widehat{\mu}_1(\widehat{\zeta}_0(\bx), \bx)$ and $\widehat{\mu}_0(\widehat{\zeta}_0(\bx), \bx)$.
    \item \textit{Estimate \( \Delta_S(\bx) \)}: Compute the residual treatment effect as:
    \begin{equation}
    \widehat{\Delta}_S(\bx) =  \widehat{\mu}_1(\widehat{\zeta}_0(\bx),\bx) -  \widehat{\mu}_0(\widehat{\zeta}_0(\bx),\bx), \label{delta_s_estimate}
\end{equation}
\end{enumerate}

\noindent \textbf{Step 5: Estimate the Proportion of Treatment Effect Explained, $R_S(\bx)$.} Using the estimates of \( \Delta(\bx) \) and \( \Delta_S(\bx) \), calculate: $ \widehat{R}_S(\bx) = 1 - \widehat{\Delta}_S(\bx)/ \widehat{\Delta}(\bx).$

We fit all learners in \texttt{R}, using the standard \texttt{lm()} function available in base \texttt{R} for the linear base learners, the \texttt{gam()} function from the \texttt{mgcv} library for the GAM learners, and the \texttt{regression\_forest()} function from the \texttt{grf} library for the regression forest learners. The \texttt{gam()} function represents the smooth functions of the specified covariates using penalized regression splines and selects the optimal basis functions for these splines via generalized cross-validation for smoothing parameter estimation \citep{wood2017generalized,Wood2023}. The function \texttt{regression\_forest()} automatically tunes several parameters to optimize the predictive performance of the forest, selecting the number of trees as 2000 to ensure a sufficiently large ensemble to reduce variance and randomly selecting the number of variables at each split as the square root of the total number of predictors in order to balance bias and variance. In addition, the function adaptively determines the minimum node size by using separate subsamples for growing and evaluating splits and optimizes the splitting rule using a generalized variance reduction criterion. Furthermore, trees are grown in a randomized way by considering only a fraction of possible splits at each node; further details can be found in \citet{Athey2019} and \citet{grf2023}. 

For variance estimation, we used the nonparametric bootstrap with 200 iterations, obtaining all $(1-\alpha)$\% confidence intervals for $\Delta(\bx)$, $\Delta_S(\bx)$, and $R_S(\bx)$ as the $\alpha/2$ and $1-\alpha/2$ percentiles of the bootstrap distributions. For the GAM and regression forest learners, all tuning parameters were held constant at the values selected in the original sample, meaning they were not re-tuned within each bootstrap iteration.

\subsection{Alternative Approaches \label{alternative}}

There are two particularly notable aspects of our algorithm within Step 4. The first is that we use the fitted learner $\widehat{\zeta}_0(\cdot)$ to predict $S^{(0)}$ given $\bX = \bx$ for every observation in the test set, including those for whom we have observed $S^{(0)}$, i.e., those in the control group. We do this to ensure consistency in estimation, mitigate noise from individual observations, and reduce potential selection bias. Specifically, using the model-based prediction $\widehat{\zeta}_0(\bx)$ rather than raw observed values of $S^{(0)}$ helps smooth out idiosyncratic variation in $S^{(0)}$ and provides a structured way to integrate information across the data. Additionally, if the distribution of \(\bX\) differs between the treated (\(G=1\)) and control (\(G=0\)) groups, as we expect it to, directly using the observed \(S^{(0)}\) in \(G=0\) may fail to capture the counterfactual distribution of \(S^{(0)}\) for \(G=1\). The model imposes a common structure that helps address this discrepancy, thereby possibly improving generalization across \(\bX\). However, one may alternatively consider using \(\widehat{\zeta}_0(\cdot)\) to predict \(S^{(0)}\) only for the treated group (\(G=1\)), while directly using the observed \(S^{(0)}\) for the control group (\(G=0\)). This approach avoids unnecessary estimation error in cases where \(S^{(0)}\) is observed without noise and the model may introduce bias. If the learner for \(\zeta_0(\bx)\) is misspecified, replacing true observations with predictions in \(G=0\) could reduce the accuracy of subsequent estimations. Thus, a reasonable diagnostic would be to compare the distributions of the observed \(S^{(0)}\) and the predicted \(\widehat{\zeta}_0(\bx)\) in the control group. If these distributions align closely, using \(\widehat{\zeta}_0(\bx)\) for everyone in both groups is unlikely to introduce significant bias. If they differ substantially, then directly using observed values for \(G=0\) may be preferable. 

The second aspect is that we use a plug-in estimator $\widehat{\mu}_g(\widehat{\zeta}_0(\bx), \bx)$ as an approximation to the  integral $\int \mu_g(s,\bx) dF_{S|\bX, G=0}(s)$.  Alternative approaches such as Monte Carlo or quadrature integration could be considered, though these approaches may suffer from increased computational cost and potential instability in high-dimensional or data-sparse regions. For example, a Monte Carlo approach would involve estimating the conditional distribution $F_{S | \bX, G=0}$ and drawing samples $\{ S_j^{(0)} \}_{j=1}^{n}$ from this distribution for a given $\bx$. The integral could then be approximated as $\frac{1}{n} \sum_{j=1}^{n} \widehat{\mu}_g(S_j^{(0)}, \bx).$ A practical way to obtain these samples nonparametrically is through weighted resampling from observed training data, using a kernel-based density estimator or a learner trained to model $F_{S | \bX, G=0}$. However, this approach becomes infeasible as the dimension of $\bX$ grows, due to the curse of dimensionality, making nearest-neighbor-based sampling unreliable. Quadrature integration, on the other hand, would require additional parametric assumptions about $F_{S | \bX, G=0}$, such as assuming a Gaussian or another well-specified parametric family for $S | \bX$, in order to determine integration points and weights. While a quadrature method can be highly accurate in a lower-dimensional setting, its effectiveness diminishes in higher dimensions unless strong assumptions on the structure of $F_{S | \bX, G=0}$ are correctly specified. Thus, while numerical integration methods would be an alternative to our proposed plug-in estimator, their practical implementation is likely infeasible as it depends on the complexity of $\bX$ and the inherent difficulty of estimating $F_{S | \bX, G=0}$.

\subsection{Asymptotic Properties \label{inference}}

In this section, we consider the statistical properties of the proposed estimators, focusing on consistency and asymptotic behavior of our three base learners: linear models, GAMs, and regression forests. For each learner, we demonstrate the consistency of $\widehat \Delta(\bx)$ and $\widehat \Delta_S(\bx)$; it will then follow from these properties that $\widehat R_S(\bx)$ is a consistent estimator of $R_S(\bx)$. 

We first discuss consistency of $\widehat \lambda_0(\bx)$ and $\widehat \lambda_1(\bx)$, for each base learner under assumptions (C1)-(C5). When these components are consistent, it follows that $\widehat \Delta(\bx)$ is a consistent estimate of $\Delta(\bx)$. When the base learners are linear models and the linear models are correctly specified, consistency of $\widehat \lambda_0(\bx)$ and $\widehat \lambda_1(\bx)$ follow from classical properties of ordinary least squares (OLS) regression. When the base learners are GAMs and the additive effects are correctly specified, consistency of the estimators as implemented here in the \texttt{mgcv} package has been shown in prior work \citep{wood2000modelling,wood2004stable,wood2011fast,wood2016smoothing} under appropriate smoothness and regularization conditions. When the base learners are regression forests, consistency of the estimators as implemented here in the \texttt{grf} package via honest trees has been shown more recently by \citet{Athey2019} and \citet{wager2018estimation}, discussed further in \citet{kunzel2019metalearners}, and \citet{caron2022estimating}.

Regarding $\widehat{\Delta}_S(\bx)$, the consistency of each component \(\mu_0(s, \bx)\), \(\mu_1(s, \bx)\), and \(\zeta_0(\bx)\) follow from the previous paragraph with the various learners. The more delicate aspect is the validity of our use of the plug-in estimator $\widehat{\mu}_g(\widehat{\zeta}_0(\bx), \bx)$ as an approximation to the integral $ \int \mu_g(s,\bx) dF_{S|\bX, G=0}(s)$, ensuring that $\widehat{\mu}_g(\widehat{\zeta}_0(\bx), \bx)$ is a consistent estimator of the integral. This approximation is valid under (C5), which ensures that small perturbations in $s$ lead to controlled deviations in $\mu_g(s, \bx)$ and therefore that $\mu_g(\zeta_0(\bx), \bx)$ is a good approximation to $E(\mu_g(S, \bx) | \bX)$ \citep{newey1994large}, and when:  

\begin{enumerate} 
\item[] (C6) $\var(S\supzero \mid \bX)$ is sufficiently small such that $\zeta_0(\bx) = E(S\supzero \mid \bX)$ is representative of $S \supzero$.
\end{enumerate} \vspace{-2mm}

\noindent Essentially, Condition (C6) allows us to claim that we minimize the approximation error in the integral via a second-order Taylor expansion, as the expectation smooths out higher-order terms \citep{vandervaart1998asymptotic}. Under these conditions, the error in replacing the integral with the plug-in estimator vanishes asymptotically, ensuring that $\widehat{\Delta}_S(\bx)$ is a consistent estimator of $\Delta_S(\bx)$. If these conditions do not hold, one may instead consider the use of numerical integration methods described in Section \ref{alternative}.

\section{Individual Identification}\label{identify}

In the previous section, we introduced several meta-learners to estimate $R_S(\bx)$, which quantifies the strength of the surrogate as a function of $\bx$. Here, we leverage these estimates to identify individuals for whom the surrogate is sufficiently strong to replace the primary outcome. After all, the ultimate goal of investigating surrogacy and heterogeneity is to guide the effective and appropriate use of surrogate markers in a future study.

Recall that higher values of $R_S(\bx)$ indicate stronger surrogacy. Though there is no established threshold for what value reflects a valid surrogate, previous work has often considered a surrogate to be ``strong" if this value or the lower bound of its confidence interval exceeds 0.50 or 0.75 \citep{bycott1998evaluation, lin1997estimating}. We denote this threshold as $\kappa$. Ideally, $\kappa$ should be selected \textit{a priori}, informed by domain expertise and study-specific considerations. However, it may also be treated as a tunable parameter when factoring in cost-effectiveness, a topic we discuss further in Section \ref{discussion}.
 
Given a chosen $\kappa$ and a specific  $\bx_i$, our goal is to determine whether the surrogate is sufficiently strong by testing the following null hypothesis:
$$H_0: R_S(\bx_i) \leq \kappa$$
for an individual $i$. We consider testing $H_0$ by constructing a one-sided $(1-\alpha)$\% confidence interval for $R_S(\bx_i)$ using our bootstrap samples and rejecting $H_0$ if $\kappa$ is less than the lower bound of the interval. More specifically, we calculate
$$p_i = \frac{1}{B} \sum_{i=1}^B I(R_S^{(b)}(\bx_i) \leq \kappa),$$
where $R_S^{(b)}(\bx_i)$ is the bootstrapped estimate of $R_S(\bx_i)$ from the $b$-th bootstrap sample, $B$ is the number of bootstrapping iterations ($B=200$ in our simulation study), and $I$ is the indicator function. To account for multiple testing, we additionally apply the Benjamini-Hochberg procedure to all calculated $p_i$'s and conclude that the surrogate is sufficiently strong for individual $\bx_i$ if the adjusted $p_i$ is less than $\alpha$ \citep{benjamini1995controlling}.

We investigate the performance of this individualized identification approach in Section \ref{sims}, in settings where the true PTE is known, by examining the positive predictive value (PPV), negative predictive value (NPV), specificity, and sensitivity of the testing results.

\section{Simulation Study} \label{sims}
\subsection{Simulation Goals and Setup}

We conducted a simulation study to evaluate the performance of our proposed methods across multiple settings, each designed to examine the tradeoffs between simple versus more complex base learners. Specifically, we considered three primary settings of increasing complexity in their data generating processes. Setting 1 featured a linear data generating process, with the true PTE, $R_S(\bx)$, ranging from 0.32 to 0.65, and where linear models were expected to perform well. Setting 2 introduced nonlinear components to the data generating process but remained additive in nature ($R_S(\bx)$ ranging from 0.14 to 0.64), making it particularly suitable for GAM base learners. Setting 3 incorporated more complex relationships ($R_S(\bx)$ ranging from 0.14 to 0.64) that violated the additive assumption of GAMs. To reflect a real-world (non-randomized) setting, in all simulation settings, the treatment assignment $G$ was dependent on the baseline covariates and was constructed such that the treatment group sizes were approximately equal. Details of the simulation settings are provided in Appendix A, along with an additional Setting 4 featuring no heterogeneity (that is, $R_S(\bx) = 0.67$ for all $\bx$). 

All settings had a sample size of $n=2000$, a test set of 200 randomly selected individuals, and six baseline covariates comprising $\bX$. In Settings 1-3, there was heterogeneity in surrogacy with respect to the first covariate, $X_1$; PTE was constant with respect to the other baseline covariates. Bootstrapped estimates with 200 iterations were used for standard error estimation and confidence interval construction. For the purpose of individual identification as described in Section \ref{identify}, we used a threshold of $\kappa = 0.5$. All simulation results were summarized across 1000 iterations, and performance was summarized in terms of median absolute bias, empirical standard error (ESE) in terms of the median absolute deviation, median standard error (ASE), median squared error, and confidence interval coverage of the true $R_S(\bx)$. 

\subsection{Simulation Results}

\begin{figure}
    \centering 
    \includegraphics[scale = 0.1]{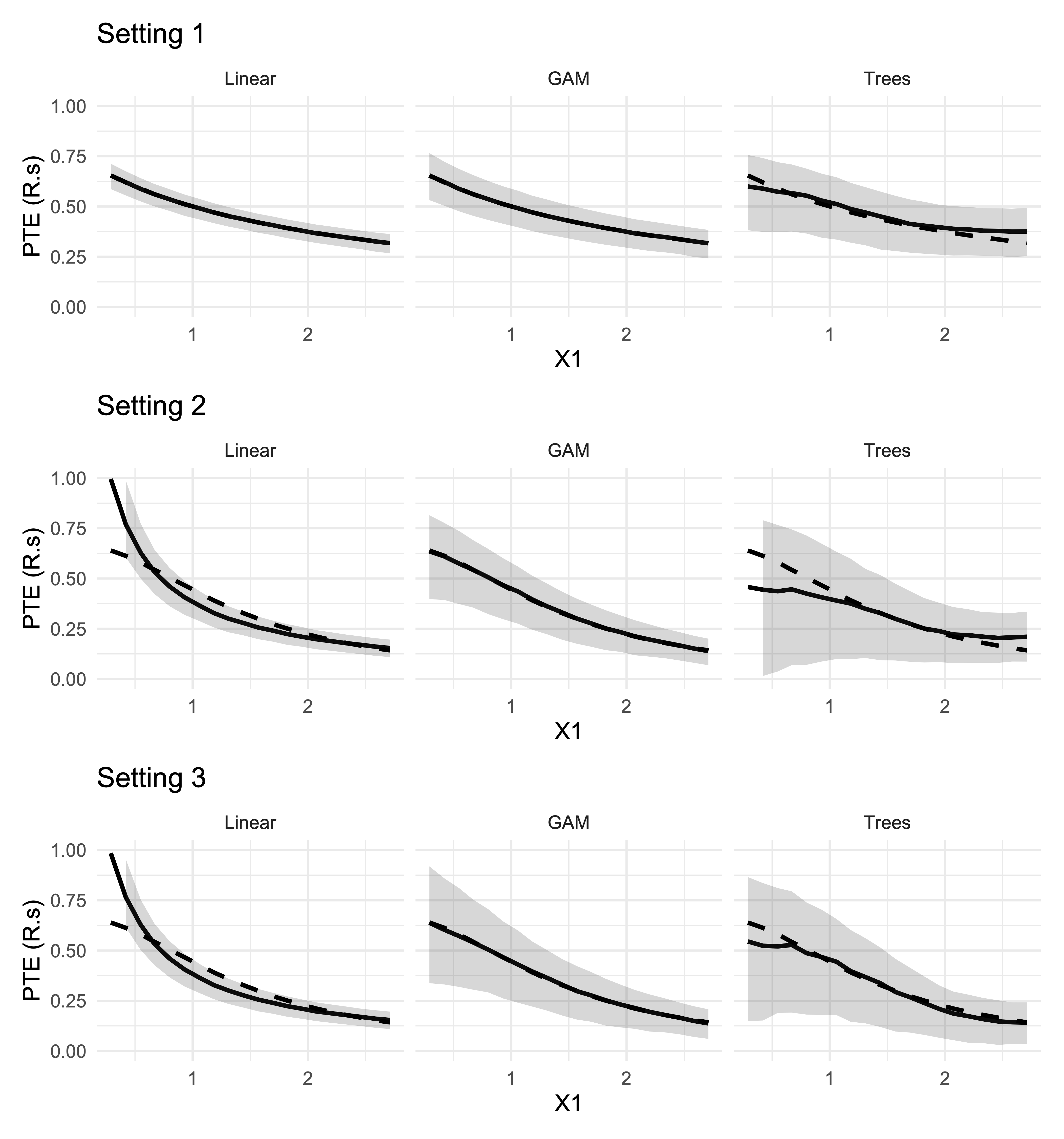}
    \caption{
    Estimated $R_S(\bx)$ (solid lines) vs. true $R_S(\bx)$ (dashed lines) plotted against $X_1$, the baseline covariate featuring heterogeneous surrogate strength in our simulations, with pointwise confidence bands (grey shading) obtained using bootstrapping.
    } \label{sim_plot}
\end{figure}

Figure \ref{sim_plot} displays the resulting estimates (solid line) and confidence intervals (gray shading) for $R_S(\bx)$, plotted against the truth (dashed line), for Settings 1-3 featuring linear, GAM, and regression forest base learners. The figure shows that the approach using linear base learners perform exceptionally well in Setting 1 as expected, with very little bias and low variance compared to the more complex learners. However, the linear base learners produce biased results for some ranges of $X_1$ in Settings 2 and 3, when the true data generating process is not linear. Interestingly, the GAM base learners perform quite well not only in Setting 2 (which we would expect), but also in Setting 3 when the data generating process was not additive. The regression forests perform reasonably well across settings, but can have some volatility in the estimates since they do not impose as much of a structure as the linear and GAM models, and the regression forests tend to have a higher variance in estimates. 

The overall results of our simulation settings are summarized numerically in Table \ref{simres}. These results are averaged over the grid of $X_1$. Again, Setting 1 showcases the strong performance of the linear base learners when appropriate, with low bias, small standard errors, and coverage reasonably close to the nominal 95\% confidence level. In Settings 2 and 3, when the linear models do not hold, coverage deteriorates due to the estimates being biased in some regions of the covariate space. Meanwhile, the GAM base learners continue to perform well in terms of high coverage and low MSE. Even though the absolute bias is somewhat higher in Setting 3 when the assumptions of the GAM are violated, the model still performs quite well overall. Across settings, the regression forests are less well-behaved, which is a known feature of regression forests without a very large sample size. Even so, the trees perform reasonably well in terms of coverage levels close to 95\% and small MSE, with higher standard errors as expected. Throughout settings and choice of base learners, the ASE estimated via resampling is reasonably close to the ESE. 

To evaluate the individual identification procedure described in Section \ref{identify}, we use our proposed approach to identify people in the testing set as those for whom the surrogate is strong ($\widehat R_S(\bx_i) > \kappa$), and we compare to the true status ($R_S(\bx_i)>\kappa)$, where $\kappa = 0.5$ in all settings. The performance is summarized in terms of positive predictive value (PPV), negative predictive value (NPV), specificity, and sensitivity in Table \ref{IDRegion_res}. Across all settings and choices of base learners, we see strong performance in terms of PPV and specificity ranging from 0.783-1 (note that higher values indicate better performance for all four quantities). The NPV is reasonably high, ranging from 0.675-0.84 across settings and learners, but lower for the tree base learners. In contrast, sensitivity is quite low, particularly for the tree base learners, meaning that among individuals where the surrogate is truly strong, the methods struggle to correctly identify a high proportion of them as such. For the linear base learners, it is worth noting (as seen in Figure \ref{sim_plot}) that the bias in Settings 2 and 3 is a result of estimating the PTE to be \textit{higher} than the truth, and thus even in these settings, the linear base learners are reasonably successful at individual identification. Of course, this is particular to this simulation setting and is not expected to hold generally for linear base learners. Compared to the linear base learners, sensitivity is lower for the GAM and tree-based approaches, suggesting that the identification procedure can be quite conservative. Notably, in practice, it is likely preferable to be more conservative in identifying individuals for whom it is appropriate to substitute the surrogate rather than less conservative.

Overall, these results demonstrate reasonable performance of the proposed methods in various settings using different base learners in terms of both estimation of $R_S(\bx)$ and individual identification. \texttt{R} code to reproduce these simulation results is available at: \url{https://github.com/rebeccaknowlton/obshetsurr-simulations}.

\section{Example} \label{example}
We illustrate our proposed framework using data from the National Health and Nutrition Examination Survey (NHANES), which is a routine national survey administered by the United States Centers for Disease Control and Prevention (CDC) National Center for Health Statistics \citep{nhanes2021}. NHANES aims to measure the health and nutrition of adults and children in the United States and includes health exams and laboratory work. 

We focus on examining the difference in fasting plasma glucose levels between obese and non-obese individuals, where obesity is defined as a body mass index (BMI) of 30 or greater. In this context, fasting plasma glucose serves as the primary outcome of interest, while the treatment/exposure is obesity status, classified as obese (treated) versus non-obese (control). Numerous studies have established a strong association between obesity and elevated fasting plasma glucose levels, a critical indicator of metabolic health and a key risk factor for serious health conditions \citep{lazar2005obesity,chandrasekaran2024role, gerstein1997glucose,emerging2011diabetes}. As a potential surrogate marker, we consider hemoglobin A1c (HbA1c), a biomarker that reflects long-term glucose regulation. Unlike fasting plasma glucose, HbA1c does not require fasting before laboratory testing, making it more convenient to measure and reducing participant burden. Our proposed framework is particularly relevant in this setting, as obesity is not randomly assigned—a key assumption underlying many traditional methods for surrogate marker evaluation. The baseline covariates, $\bX$, for this illustration are age, sex, and (total) cholesterol. Therefore, our overall goal is to use our framework to examine the potential heterogeneity (with respect to age, sex, and cholesterol) in surrogate strength when considering HbA1c as a surrogate for fasting plasma glucose when comparing obese to non-obese individuals.

\begin{landscape}
\begin{figure}
    \centering     
    \includegraphics[scale = 0.4]{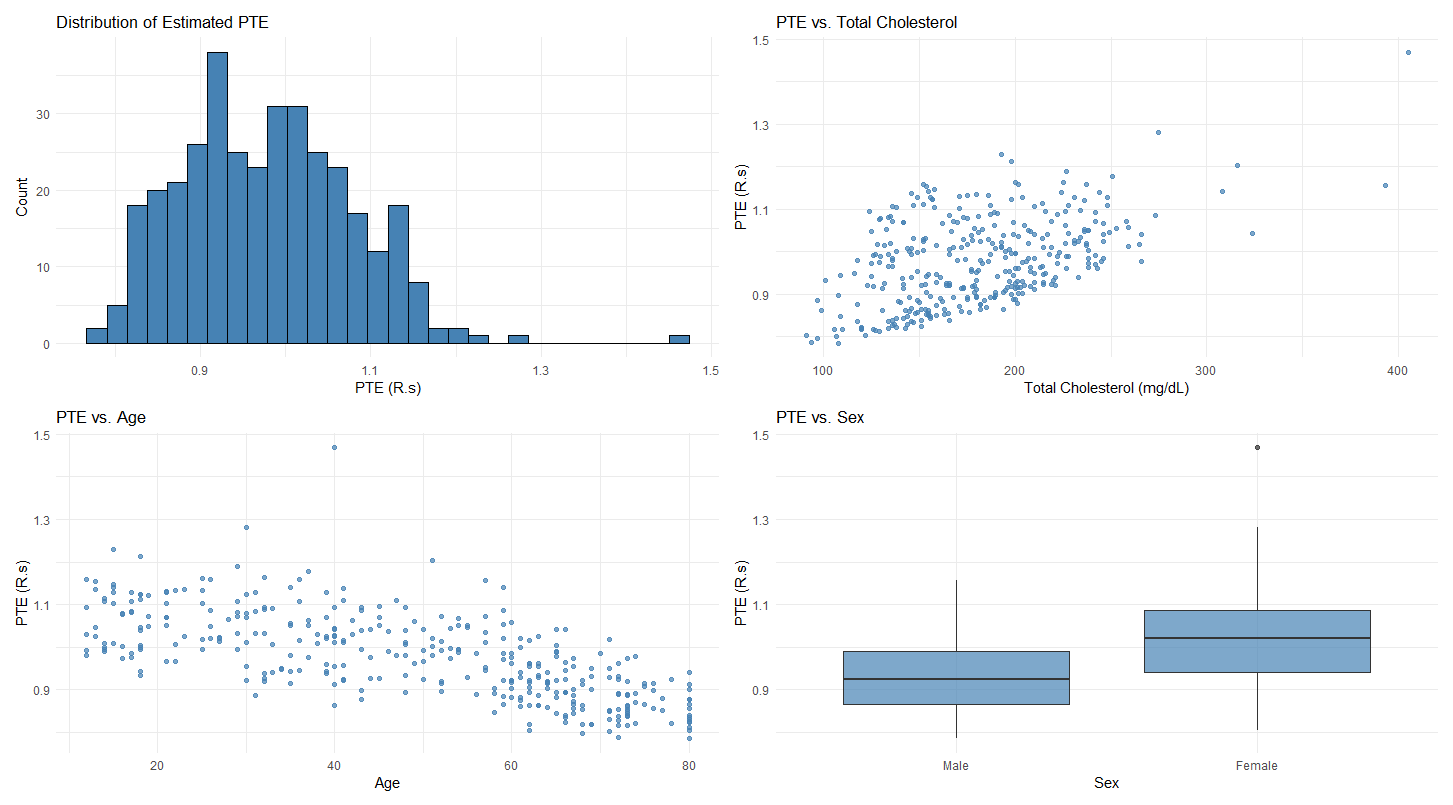}

    \caption{
    Estimation results for the NHANES survey data, evaluating the strength of HbA1c as a surrogate marker for plasma fasting glucose, when the exposure is obesity status; subfigures show the distribution of PTE estimates (top left panel) and PTE estimates by cholesterol (top right), age (bottom left), and sex (bottom right).
    } \label{nhanes_plot}
\end{figure}
\end{landscape}

We use cross-sectional survey data from the 2-year cycle August 2021--August 2023, including adults and children, which is publicly available on the CDC's website. Individuals missing fasting plasma glucose, HbA1c, BMI,  age, sex, and cholesterol were excluded. Our final analytic sample size was $n=3476$, with $n_0 = 2158$ non-obese individuals and $n_1=1318$ obese individuals. We split our sample, retaining 350 observations for testing data (about 10\% of the total sample size, similar to our numerical studies). In practice, applying our framework requires selecting appropriate base learners, which will naturally be context dependent. For this illustration, we demonstrate linear models as the base learners; in Appendix B, we additionally include results using GAMs and regression forests as the base learners. We used our approach to obtain PTE estimates for the testing data set, the results of which are displayed in Figure \ref{nhanes_plot}.  Recall that, as discussed in Section \ref{proposed}, $R_S(\bx) $ may be larger than 1. In the top left panel, we show the distribution of PTE estimates which demonstrates that the estimated PTE is generally high, i.e., HbA1c appears to be a good surrogate for plasma fasting glucose in this data set. The remaining panels show PTE estimates by cholesterol (top right), age (bottom left), and sex (bottom right), indicating that the surrogate strength varies across these different baseline characteristics, with higher estimated PTE for individuals with higher cholesterol, younger ages, and for females.

To demonstrate the practical application of our estimates for a future study, we consider six hypothetical future patients with specific baseline characteristics, as shown in Table \ref{NHANES_table}. Leveraging the trained models from the NHANES data, we compute 95\% confidence intervals for each patient's PTE, illustrating how surrogate strength varies based on individual characteristics. As a potential application in future studies, one might deem the surrogate marker a suitable substitute if the lower bound of the confidence interval is at least 0.70. Under this criterion, the surrogate alone would suffice for patients 2, 3, 4, and 6, whereas the primary outcome would still need to be measured for patients 1 and 5. These findings highlight the ability of our approach to incorporate patient-specific information, facilitating more tailored decisions about future outcome measurement in studies involving non-randomized exposures. R code to reproduce results is available at: \url{https://github.com/rebeccaknowlton/obshetsurr-NHANES-example}.

\section{Discussion} \label{discussion}

We have proposed a framework for evaluating heterogeneous surrogate strength in observational settings characterized by complex covariate relationships. Our methodology offers flexibility via different choices of base learners within the T-Learner, ranging from computationally efficient linear models to more complex tree-based algorithms. Our individual-level approach to evaluating surrogate validity aligns with the growing emphasis on personalized decision-making, especially in contexts involving complex and heterogeneous data \citep{kent2018personalized, mueller2023personalized}. Rather than relying on a rigid, one-size-fits-all decision rule, our framework enables robust, data-driven decisions tailored to specific individual characteristics. In addition, we developed appropriate statistical tests for evaluating surrogate strength as measured using clinically relevant thresholds and validated the performance of our methods through simulation studies. An \texttt{R} package implementing our methods, \texttt{cohetsurr}, is available on CRAN \citep{cohetsurr}.

Our framework notably shares mathematical similarities with the topic of moderated mediation; however, there are fundamental conceptual differences in the objectives that are worth discussing. Moderated mediation focuses on understanding the mechanism through which a treatment affects an outcome and how this mechanism varies across subgroups \citep{qin2023, li2023effect}. In contrast, our approach to surrogate markers is not necessarily concerned with establishing causal mechanisms, but rather with identifying variables that reliably capture the treatment effect and thus can substitute for the primary outcome in future studies. This distinction is crucial: while mediation analysis seeks to decompose and explain causal pathways, surrogate evaluation in the PTE framework aims to validate replacement outcomes that capture treatment effects, regardless of the underlying mechanisms. Recent methodological advances in heterogeneous mediation effects, such as the Bayesian tree ensemble approach in \cite{ting2023}, share our interest in effect heterogeneity but differ fundamentally in their goal of understanding mechanistic pathways rather than outcome substitution. Our framework therefore complements rather than overlaps with these developments. 

Notably, our approach relies on strong, untestable causal assumptions that, while common in the literature, may not hold in practice. Furthermore, more assumptions may be needed if one is interested in ensuring $R_S(\bx) \in [0,1]$ and guarding against the surrogate paradox---a phenomenon where a positive treatment effect on the surrogate and positive surrogate-outcome association paradoxically coexist with a negative treatment effect on the primary outcome. Protection against this paradox typically requires additional assumptions: monotonicity in the surrogate-outcome relationship, a non-negative treatment effect on the surrogate, and non-negative direct treatment effects conditional on the surrogate and baseline characteristics \citep{vanderweele2013, chen2007criteria,hsiao2025}. While these conditions are important when using surrogates as outcome replacements in future studies, they are less critical in our context where we focus on evaluating surrogate strength in a single study where the primary outcome is also observed. Still, researchers applying our methods should consider whether such additional assumptions might be warranted for their specific application, particularly if the findings will inform future surrogate-based studies.

With respect to the statistical properties of our proposed methods, it is important to consider the convergence rates of the various estimators. The T-Learner approach, while offering flexibility, faces inherent challenges in estimation efficiency and, in fact, often perform poorly when the true heterogeneity is simple, or when the treatment groups are very different sizes \citep{kunzel2019metalearners, caron2022estimating}. By fitting separate models for the treatment and control groups, we effectively reduce the available sample size for each model, potentially slowing convergence even when we have consistency. This challenge is further compounded by our sample-splitting procedure. The convergence concerns are especially pronounced when using  machine learning methods like regression trees, which typically require substantial data to achieve reliable estimates. While simpler base learners like linear models offer faster convergence rates under limited data scenarios, they may be biased when the true underlying relationships are complex. This creates a practical trade-off between bias and variance that must be carefully considered when choosing a base learner, depending on the sample size and complexity of the data.

Our framework enables individualized surrogate identification in real-world settings, but key questions remain about leveraging this information to achieve cost savings---one of the common motivations for evaluating surrogate markers \citep{tao2017efficient, pryseley2010using, kosorok1993using}. One key consideration is the choice of $\kappa$ for the identification procedure in Section \ref{identify}, which sets the threshold for deeming a surrogate sufficiently strong. A higher $\kappa$ ensures greater confidence in the surrogate's validity but requires more extensive primary outcome measurements in a future study, while a lower $\kappa$ may reduce costs at the expense of certainty. By carefully selecting $\kappa$, researchers can balance the trade-off between cost-effectiveness and statistical confidence. To implement these cost savings in practice at a chosen threshold $\kappa$, one can consider extending recent work by \cite{knowlton2025efficient}, which developed efficient testing procedures integrating surrogate and primary outcome information from disjoint population subsets in experimental settings. Extending these methods to observational settings, with appropriate consideration of confounding and selection bias, could enable more efficient study designs that strategically combine surrogate measurements with primary outcomes across heterogeneous populations. Such developments would further enhance the practical value of heterogeneous surrogate evaluation, particularly in settings where outcome measurement is expensive or impractical. 

\section*{Acknowledgements}
This work was supported by NIDDK grant R01DK118354 (PI:Parast).

\clearpage 

\clearpage
\begin{table}[hptb]
\caption{Simulation results for $R_S(\bx)$ in Settings 1-3, averaged over $X_1$, where Bias reflects the absolute value of the difference between the estimate and the truth, summarized as the median of 1000 iterations; ESE represents the empirical standard error (calculated as the median absolute deviation of estimates across iterations); ASE represents the average standard error (calculated as the median of the bootstrap variance estimates);  MSE represents the median squared error; Coverage indicates the coverage rate of 95\% bootstrap confidence intervals with respect to the truth. \label{simres}}

\begin{center}
\begin{tabular}{|l|c|c|c|c|c|} \hline
&\multicolumn{3}{c|}{Setting 1}\\ \hline
\multicolumn{1}{|l|}{}&\multicolumn{1}{c|}{Linear}&\multicolumn{1}{c|}{GAM}&\multicolumn{1}{c|}{Trees}\\ \hline 
Bias&0.015&0.021&0.052\\ 
ESE&0.023&0.032&0.070\\ 
ASE&0.026&0.044&0.074\\ 
MSE&0.000&0.000&0.003\\ 
Coverage&0.966&0.980&0.940\\ 
\hline
&\multicolumn{3}{c|}{Setting 2}\\ \hline
\multicolumn{1}{|l|}{}&\multicolumn{1}{c|}{Linear}&\multicolumn{1}{c|}{GAM}&\multicolumn{1}{c|}{Trees}\\ \hline 
Bias&0.058&0.034&0.087\\ 
ESE&0.038&0.051&0.102\\ 
ASE&0.043&0.061&0.110\\ 
MSE&0.009&0.001&0.010\\ 
Coverage&0.764&0.967&0.919\\ 
\hline
&\multicolumn{3}{c|}{Setting 3}\\ \hline
\multicolumn{1}{|l|}{}&\multicolumn{1}{c|}{Linear}&\multicolumn{1}{c|}{GAM}&\multicolumn{1}{c|}{Trees}\\ \hline 
Bias&0.057&0.042&0.071\\ 
ESE&0.035&0.062&0.101\\ 
ASE&0.040&0.075&0.099\\ 
MSE&0.009&0.002&0.006\\ 
Coverage&0.754&0.967&0.942\\ 
\hline
\end{tabular}
\vspace{3mm}
\end{center}
\end{table}

\clearpage
\begin{table}[hptb]
\caption{Performance assessment for the individual identification procedure in Settings 1-3 and summarized over 1000 iterations. PPV reflects the positive predictive value, i.e., the proportion of people identified by our procedure as those for whom the surrogate is strong, where the surrogate is truly strong; NPV reflects the negative predictive value, i.e., the proportion of people identified by our procedure as those for whom the surrogate is weak, where the surrogate is truly weak; Specificity reflects the proportion of people for whom the surrogate is truly weak ($R_S(\bx) <= \kappa$) who have been correctly identified as such; Sensitivity reflects the proportion of people for whom the surrogate is truly strong ($R_S(\bx) > \kappa$) who have been correctly identified.   \label{IDRegion_res}}

\begin{center}
\begin{tabular}{|l|c|c|c|c|c|} \hline
&\multicolumn{3}{c|}{Setting 1}\\ \hline
\multicolumn{1}{|l|}{}&\multicolumn{1}{c|}{Linear}&\multicolumn{1}{c|}{GAM}&\multicolumn{1}{c|}{Trees}\\ \hline 
PPV&0.998&0.998&0.896\\ 
NPV&0.831&0.732&0.675\\  
Specificity&0.999&1.000&0.998\\ 
Sensitivity&0.591&0.267&0.038\\ 
\hline
&\multicolumn{3}{c|}{Setting 2}\\ \hline
\multicolumn{1}{|l|}{}&\multicolumn{1}{c|}{Linear}&\multicolumn{1}{c|}{GAM}&\multicolumn{1}{c|}{Trees}\\ \hline 
PPV&1.000&0.990&0.828\\ 
NPV&0.831&0.740&0.729\\ 
Specificity&1.000&1.000&1.000\\ 
Sensitivity&0.457&0.057&0.001\\ 
\hline
&\multicolumn{3}{c|}{Setting 3}\\ \hline
\multicolumn{1}{|l|}{}&\multicolumn{1}{c|}{Linear}&\multicolumn{1}{c|}{GAM}&\multicolumn{1}{c|}{Trees}\\ \hline 
PPV&1.000&0.984&0.873\\ 
NPV&0.840&0.735&0.731\\ 
Specificity&1.000&1.000&0.999\\ 
Sensitivity&0.489&0.035&0.016\\ 
\hline
\end{tabular}
\vspace{3mm}
\end{center}
\end{table}

\clearpage 

\clearpage
\begin{landscape}

\begin{table}[hptb]
\caption{Estimated 95\% confidence intervals for the PTE for six hypothetical patients, based on their age, sex, and cholesterol levels using the proposed method applied to the NHANES survey data \label{NHANES_table}}
\begin{center}
\begin{tabular}{|c|c|c|c|c|} \hline
\multicolumn{1}{|l|}{Patient ID}&\multicolumn{1}{l|}{Age}&\multicolumn{1}{l|}{Sex}&\multicolumn{1}{l|}{Cholesterol (mg/dL)}&\multicolumn{1}{l|}{Estimated 95\% Confidence Interval for PTE}\\ \hline
1~~~~~~~~~~~&65~~~~~~~~~~&Male~~~~~~~~&160~~~~~~~~~&(0.69,~1.12)\\ 
2~~~~~~~~~~~&45~~~~~~~~~~&Female~~~~~~&220~~~~~~~~~&(0.87,~1.43)\\ 
3~~~~~~~~~~~&35~~~~~~~~~~&Female~~~~~~&250~~~~~~~~~&(0.82,~2.86)\\ 
4~~~~~~~~~~~&50~~~~~~~~~~&Male~~~~~~~~&180~~~~~~~~~&(0.74,~1.17)\\ 
5~~~~~~~~~~~&70~~~~~~~~~~&Male~~~~~~~~&140~~~~~~~~~&(0.60,~1.12)\\ 
6~~~~~~~~~~~&30~~~~~~~~~~&Female~~~~~~&210~~~~~~~~~&(0.89,~1.78)\\ 
\hline
\end{tabular}
\vspace{3mm}
\end{center}
\end{table}
\end{landscape}

\clearpage
\bibliographystyle{biom}
 \bibliography{bib}

\begin{thebibliography}{}

\bibitem[\protect\citeauthoryear{Agniel, Hejblum, Thi{\'e}baut, and
  Parast}{Agniel et~al.}{2023}]{agniel2023doubly}
Agniel, D., Hejblum, B.~P., Thi{\'e}baut, R., and Parast, L. (2023).
\newblock Doubly robust evaluation of high-dimensional surrogate markers.
\newblock {\em Biostatistics} {\bf 24,} 985--999.

\bibitem[\protect\citeauthoryear{Agniel and Parast}{Agniel and
  Parast}{2024}]{agniel2024robust}
Agniel, D. and Parast, L. (2024).
\newblock Robust evaluation of longitudinal surrogate markers with censored
  data.
\newblock {\em Journal of the Royal Statistical Society Series B: Statistical
  Methodology} {\bf qkae119,}.

\bibitem[\protect\citeauthoryear{Andrews and Didelez}{Andrews and
  Didelez}{2021}]{andrews2021insights}
Andrews, R.~M. and Didelez, V. (2021).
\newblock Insights into the cross-world independence assumption of causal
  mediation analysis.
\newblock {\em Epidemiology} {\bf 32,} 209--219.

\bibitem[\protect\citeauthoryear{Athey and Imbens}{Athey and
  Imbens}{2016}]{athey2016recursive}
Athey, S. and Imbens, G. (2016).
\newblock Recursive partitioning for heterogeneous causal effects.
\newblock {\em Proceedings of the National Academy of Sciences} {\bf 113,}
  7353--7360.

\bibitem[\protect\citeauthoryear{Athey, Tibshirani, and Wager}{Athey
  et~al.}{2019}]{Athey2019}
Athey, S., Tibshirani, J., and Wager, S. (2019).
\newblock Generalized random forests.
\newblock {\em Annals of Statistics} {\bf 47,} 1148--1178.

\bibitem[\protect\citeauthoryear{Athey and Wager}{Athey and
  Wager}{2019}]{athey2019estimating}
Athey, S. and Wager, S. (2019).
\newblock Estimating treatment effects with causal forests: An application.
\newblock {\em Observational studies} {\bf 5,} 37--51.

\bibitem[\protect\citeauthoryear{Benjamini and Hochberg}{Benjamini and
  Hochberg}{1995}]{benjamini1995controlling}
Benjamini, Y. and Hochberg, Y. (1995).
\newblock Controlling the false discovery rate: a practical and powerful
  approach to multiple testing.
\newblock {\em Journal of the Royal Statistical Society: Series B
  (Methodological)} {\bf 57,} 289--300.

\bibitem[\protect\citeauthoryear{Boyko}{Boyko}{2013}]{boyko2013observational}
Boyko, E.~J. (2013).
\newblock Observational research—opportunities and limitations.
\newblock {\em Journal of Diabetes and its Complications} {\bf 27,} 642--648.

\bibitem[\protect\citeauthoryear{Bycott and Taylor}{Bycott and
  Taylor}{1998}]{bycott1998evaluation}
Bycott, P.~W. and Taylor, J.~M. (1998).
\newblock An evaluation of a measure of the proportion of the treatment effect
  explained by a surrogate marker.
\newblock {\em Controlled Clinical Trials} {\bf 19,} 555--568.

\bibitem[\protect\citeauthoryear{Caron, Baio, and Manolopoulou}{Caron
  et~al.}{2022}]{caron2022estimating}
Caron, A., Baio, G., and Manolopoulou, I. (2022).
\newblock Estimating individual treatment effects using non-parametric
  regression models: A review.
\newblock {\em Journal of the Royal Statistical Society Series A: Statistics in
  Society} {\bf 185,} 1115--1149.

\bibitem[\protect\citeauthoryear{{C{D}{C}}}{{C{D}{C}}}{2025}]{nhanes2021}
{C{D}{C}} (2025).
\newblock United states centers for disease control and prevention national
  health and nutrition examination survey data ({A}ugust 2021--{A}ugust 2023).
\newblock Accessed: 2025-02-05.

\bibitem[\protect\citeauthoryear{Chandrasekaran and Weiskirchen}{Chandrasekaran
  and Weiskirchen}{2024}]{chandrasekaran2024role}
Chandrasekaran, P. and Weiskirchen, R. (2024).
\newblock The role of obesity in type 2 diabetes mellitus—an overview.
\newblock {\em International Journal of Molecular Sciences} {\bf 25,} 1882.

\bibitem[\protect\citeauthoryear{Chen, Geng, and Jia}{Chen
  et~al.}{2007}]{chen2007criteria}
Chen, H., Geng, Z., and Jia, J. (2007).
\newblock Criteria for surrogate end points.
\newblock {\em Journal of the Royal Statistical Society Series B: Statistical
  Methodology} {\bf 69,} 919--932.

\bibitem[\protect\citeauthoryear{Collaboration}{Collaboration}{2011}]{emerging2011diabetes}
Collaboration, E. R.~F. (2011).
\newblock Diabetes mellitus, fasting glucose, and risk of cause-specific death.
\newblock {\em New England Journal of Medicine} {\bf 364,} 829--841.

\bibitem[\protect\citeauthoryear{Fleming}{Fleming}{1994}]{fleming1994}
Fleming, T.~R. (1994).
\newblock Surrogate markers in {AIDS} and cancer trials.
\newblock {\em Statistics in Medicine} {\bf 13,} 1423--1435.

\bibitem[\protect\citeauthoryear{Freedman, Graubard, and Schatzkin}{Freedman
  et~al.}{1992}]{freedman1992}
Freedman, L.~S., Graubard, B.~I., and Schatzkin, A. (1992).
\newblock Statistical validation of intermediate endpoints for chronic
  diseases.
\newblock {\em Statistics in medicine} {\bf 11,} 167--178.

\bibitem[\protect\citeauthoryear{Gerstein}{Gerstein}{1997}]{gerstein1997glucose}
Gerstein, H. (1997).
\newblock Glucose: a continuous risk factor for cardiovascular disease.
\newblock {\em Diabetic Medicine} {\bf 14,} S25--S31.

\bibitem[\protect\citeauthoryear{Han, Wang, and Cai}{Han
  et~al.}{2022}]{han2022identifying}
Han, L., Wang, X., and Cai, T. (2022).
\newblock Identifying surrogate markers in real-world comparative effectiveness
  research.
\newblock {\em Statistics in Medicine} {\bf 41,} 5290--5304.

\bibitem[\protect\citeauthoryear{Hsiao, Tian, and Parast}{Hsiao
  et~al.}{2025}]{hsiao2025}
Hsiao, E., Tian, L., and Parast, L. (2025).
\newblock Avoiding the surrogate paradox: An empirical framework for assessing
  assumptions.
\newblock {\em Journal of Nonparametric Statistics, In press} .

\bibitem[\protect\citeauthoryear{Imai, Keele, and Tingley}{Imai
  et~al.}{2010}]{imai2010general}
Imai, K., Keele, L., and Tingley, D. (2010).
\newblock A general approach to causal mediation analysis.
\newblock {\em Psychological methods} {\bf 15,} 309.

\bibitem[\protect\citeauthoryear{Katz}{Katz}{2004}]{katz2004}
Katz, R. (2004).
\newblock Biomarkers and surrogate markers: an fda perspective.
\newblock {\em NeuroRx} {\bf 1,} 189--195.

\bibitem[\protect\citeauthoryear{Kent, Steyerberg, and Van~Klaveren}{Kent
  et~al.}{2018}]{kent2018personalized}
Kent, D.~M., Steyerberg, E., and Van~Klaveren, D. (2018).
\newblock Personalized evidence based medicine: predictive approaches to
  heterogeneous treatment effects.
\newblock {\em BMJ} {\bf 363,}.

\bibitem[\protect\citeauthoryear{Knowlton}{Knowlton}{2025}]{cohetsurr}
Knowlton, R. (2025).
\newblock {\em cohetsurr: Assessing Complex Heterogeneity in Surrogacy}.
\newblock R package version 2.0.

\bibitem[\protect\citeauthoryear{Knowlton and Parast}{Knowlton and
  Parast}{2025}]{knowlton2025efficient}
Knowlton, R. and Parast, L. (2025).
\newblock Efficient testing using surrogate information.
\newblock {\em Under Review} .

\bibitem[\protect\citeauthoryear{Knowlton, Tian, and Parast}{Knowlton
  et~al.}{2025}]{knowlton2024}
Knowlton, R., Tian, L., and Parast, L. (2025).
\newblock A general framework to assess complex heterogeneity in the strength
  of a surrogate marker.
\newblock {\em Statistics in Medicine} {\bf 44,} e70001.

\bibitem[\protect\citeauthoryear{Kosorok and Fleming}{Kosorok and
  Fleming}{1993}]{kosorok1993using}
Kosorok, M.~R. and Fleming, T.~R. (1993).
\newblock Using surrogate failure time data to increase cost effectiveness in
  clinical trials.
\newblock {\em Biometrika} {\bf 80,} 823--833.

\bibitem[\protect\citeauthoryear{K{\"u}nzel, Sekhon, Bickel, and Yu}{K{\"u}nzel
  et~al.}{2019}]{kunzel2019metalearners}
K{\"u}nzel, S.~R., Sekhon, J.~S., Bickel, P.~J., and Yu, B. (2019).
\newblock Metalearners for estimating heterogeneous treatment effects using
  machine learning.
\newblock {\em Proceedings of the National Academy of Sciences} {\bf 116,}
  4156--4165.

\bibitem[\protect\citeauthoryear{Lazar}{Lazar}{2005}]{lazar2005obesity}
Lazar, M.~A. (2005).
\newblock How obesity causes diabetes: not a tall tale.
\newblock {\em Science} {\bf 307,} 373--375.

\bibitem[\protect\citeauthoryear{Li, Mathur, Solomon, Ridker, Glynn, and
  Yoshida}{Li et~al.}{2023}]{li2023effect}
Li, Y., Mathur, M.~B., Solomon, D.~H., Ridker, P.~M., Glynn, R.~J., and
  Yoshida, K. (2023).
\newblock Effect measure modification by covariates in mediation: extending
  regression-based causal mediation analysis.
\newblock {\em Epidemiology} {\bf 34,} 661--672.

\bibitem[\protect\citeauthoryear{Lin, Fleming, and De~Gruttola}{Lin
  et~al.}{1997}]{lin1997estimating}
Lin, D., Fleming, T., and De~Gruttola, V. (1997).
\newblock Estimating the proportion of treatment effect explained by a
  surrogate marker.
\newblock {\em Statistics in Medicine} {\bf 16,} 1515--1527.

\bibitem[\protect\citeauthoryear{Meinshausen}{Meinshausen}{2007}]{meinshausen2007relaxed}
Meinshausen, N. (2007).
\newblock Relaxed lasso.
\newblock {\em Computational Statistics \& Data Analysis} {\bf 52,} 374--393.

\bibitem[\protect\citeauthoryear{Mueller and Pearl}{Mueller and
  Pearl}{2023}]{mueller2023personalized}
Mueller, S. and Pearl, J. (2023).
\newblock Personalized decision making--a conceptual introduction.
\newblock {\em Journal of Causal Inference} {\bf 11,} 20220050.

\bibitem[\protect\citeauthoryear{Newey and McFadden}{Newey and
  McFadden}{1994}]{newey1994large}
Newey, W.~K. and McFadden, D. (1994).
\newblock Large sample estimation and hypothesis testing.
\newblock {\em Handbook of Econometrics} {\bf 4,} 2111--2245.

\bibitem[\protect\citeauthoryear{Parast, Cai, and Tian}{Parast
  et~al.}{2023a}]{parast2021hetero}
Parast, L., Cai, T., and Tian, L. (2023a).
\newblock Testing for heterogeneity in the utility of a surrogate marker.
\newblock {\em Biometrics} {\bf 79,} 799--810.

\bibitem[\protect\citeauthoryear{Parast, Cai, and Tian}{Parast
  et~al.}{2023b}]{parast2022using}
Parast, L., Cai, T., and Tian, L. (2023b).
\newblock Using a surrogate with heterogeneous utility to test for a treatment
  effect.
\newblock {\em Statistics in Medicine} {\bf 42,} 68--88.

\bibitem[\protect\citeauthoryear{Preacher and Kelley}{Preacher and
  Kelley}{2011}]{preacher2011effect}
Preacher, K.~J. and Kelley, K. (2011).
\newblock Effect size measures for mediation models: quantitative strategies
  for communicating indirect effects.
\newblock {\em Psychological methods} {\bf 16,} 93.

\bibitem[\protect\citeauthoryear{Pryseley, Tilahun, Alonso, and
  Molenberghs}{Pryseley et~al.}{2010}]{pryseley2010using}
Pryseley, A., Tilahun, A., Alonso, A., and Molenberghs, G. (2010).
\newblock Using earlier measures in a longitudinal sequence as a potential
  surrogate for a later one.
\newblock {\em Computational statistics \& data analysis} {\bf 54,} 1342--1354.

\bibitem[\protect\citeauthoryear{Qin and Wang}{Qin and Wang}{2023}]{qin2023}
Qin, X. and Wang, L. (2023).
\newblock Causal moderated mediation analysis: Methods and software.
\newblock {\em Behavior Research Methods} pages 1--21.

\bibitem[\protect\citeauthoryear{Roberts, Elliott, and Taylor}{Roberts
  et~al.}{2021}]{roberts2021incorporating}
Roberts, E.~K., Elliott, M.~R., and Taylor, J.~M. (2021).
\newblock Incorporating baseline covariates to validate surrogate endpoints
  with a constant biomarker under control arm.
\newblock {\em Statistics in Medicine} {\bf 40,} 6605--6618.

\bibitem[\protect\citeauthoryear{Rosenbaum}{Rosenbaum}{2005}]{rosenbaum2005observational}
Rosenbaum, P.~R. (2005).
\newblock Observational study.
\newblock {\em Encyclopedia of statistics in behavioral science} {\bf 3,}
  1451--1462.

\bibitem[\protect\citeauthoryear{Stijven, Alonso, and Molenberghs}{Stijven
  et~al.}{2024}]{stijven2024proportion}
Stijven, F., Alonso, A., and Molenberghs, G. (2024).
\newblock Proportion of treatment effect explained: An overview of
  interpretations.
\newblock {\em Statistical Methods in Medical Research} {\bf 33,} 1278--1296.

\bibitem[\protect\citeauthoryear{Tao, Zeng, and Lin}{Tao
  et~al.}{2017}]{tao2017efficient}
Tao, R., Zeng, D., and Lin, D.-Y. (2017).
\newblock Efficient semiparametric inference under two-phase sampling, with
  applications to genetic association studies.
\newblock {\em Journal of the American Statistical Association} {\bf 112,}
  1468--1476.

\bibitem[\protect\citeauthoryear{Tibshirani, Athey, Friedberg, Hadad,
  Hirshberg, Wager, and Zhou}{Tibshirani et~al.}{2023}]{grf2023}
Tibshirani, J., Athey, S., Friedberg, R., Hadad, V., Hirshberg, D., Wager, S.,
  and Zhou, E. (2023).
\newblock {\em grf: Generalized Random Forests}.
\newblock R package version 2.3.0.

\bibitem[\protect\citeauthoryear{Ting and Linero}{Ting and
  Linero}{2023}]{ting2023}
Ting, A. and Linero, A.~R. (2023).
\newblock Estimating heterogeneous causal mediation effects with bayesian
  decision tree ensembles.
\newblock {\em arXiv preprint arXiv:2303.01620} .

\bibitem[\protect\citeauthoryear{Van~der Laan, Polley, and Hubbard}{Van~der
  Laan et~al.}{2007}]{van2007super}
Van~der Laan, M.~J., Polley, E.~C., and Hubbard, A.~E. (2007).
\newblock Super learner.
\newblock {\em Statistical Applications in Genetics and Molecular Biology} {\bf
  6,}.

\bibitem[\protect\citeauthoryear{van~der Laan and Rose}{van~der Laan and
  Rose}{2011}]{rose2011introduction}
van~der Laan, M.~J. and Rose, S. (2011).
\newblock {\em Targeted Learning}.
\newblock Springer.

\bibitem[\protect\citeauthoryear{van~der Vaart}{van~der
  Vaart}{1998}]{vandervaart1998asymptotic}
van~der Vaart, A.~W. (1998).
\newblock {\em Asymptotic Statistics}.
\newblock Cambridge University Press.

\bibitem[\protect\citeauthoryear{VanderWeele}{VanderWeele}{2013}]{vanderweele2013}
VanderWeele, T.~J. (2013).
\newblock Surrogate measures and consistent surrogates.
\newblock {\em Biometrics} {\bf 69,} 561--565.

\bibitem[\protect\citeauthoryear{Wager and Athey}{Wager and
  Athey}{2018}]{wager2018estimation}
Wager, S. and Athey, S. (2018).
\newblock Estimation and inference of heterogeneous treatment effects using
  random forests.
\newblock {\em Journal of the American Statistical Association} {\bf 113,}
  1228--1242.

\bibitem[\protect\citeauthoryear{Wang and Taylor}{Wang and
  Taylor}{2002}]{wang2002}
Wang, Y. and Taylor, J.~M. (2002).
\newblock A measure of the proportion of treatment effect explained by a
  surrogate marker.
\newblock {\em Biometrics} {\bf 58,} 803--812.

\bibitem[\protect\citeauthoryear{Wood}{Wood}{2000}]{wood2000modelling}
Wood, S.~N. (2000).
\newblock Modelling and smoothing parameter estimation with multiple quadratic
  penalties.
\newblock {\em Journal of the Royal Statistical Society: Series B (Statistical
  Methodology)} {\bf 62,} 413--428.

\bibitem[\protect\citeauthoryear{Wood}{Wood}{2004}]{wood2004stable}
Wood, S.~N. (2004).
\newblock Stable and efficient multiple smoothing parameter estimation for
  generalized additive models.
\newblock {\em Journal of the American Statistical Association} {\bf 99,}
  673--686.

\bibitem[\protect\citeauthoryear{Wood}{Wood}{2011}]{wood2011fast}
Wood, S.~N. (2011).
\newblock Fast stable restricted maximum likelihood and marginal likelihood
  estimation of semiparametric generalized linear models.
\newblock {\em Journal of the Royal Statistical Society Series B: Statistical
  Methodology} {\bf 73,} 3--36.

\bibitem[\protect\citeauthoryear{Wood}{Wood}{2017}]{wood2017generalized}
Wood, S.~N. (2017).
\newblock {\em Generalized Additive Models: An Introduction with R}.
\newblock chapman and hall/CRC.

\bibitem[\protect\citeauthoryear{Wood}{Wood}{2023}]{Wood2023}
Wood, S.~N. (2023).
\newblock {\em mgcv: Mixed GAM Computation Vehicle with Automatic Smoothness
  Estimation}.
\newblock R package version 1.8-42.

\bibitem[\protect\citeauthoryear{Wood, Pya, and S{\"a}fken}{Wood
  et~al.}{2016}]{wood2016smoothing}
Wood, S.~N., Pya, N., and S{\"a}fken, B. (2016).
\newblock Smoothing parameter and model selection for general smooth models.
\newblock {\em Journal of the American Statistical Association} {\bf 111,}
  1548--1563.

\end{thebibliography}

\clearpage

\appendix
\setcounter{table}{0}
\renewcommand{\thetable}{A\arabic{table}}
\setcounter{figure}{0}
\renewcommand{\thefigure}{A\arabic{figure}}
\renewcommand{\theequation}{A.\arabic{equation}}

\allowdisplaybreaks

\section*{Appendix A} 

Here, we describe in detail the data generating process for our simulation settings. In addition to Settings 1-3 described in the main text, here we include an additional Setting 4 that features \textit{no} heterogeneity in terms of PTE, as a control setting.

In all settings, the baseline covariates are $X_1 \sim U(0,3)$, $X_2 \sim Gamma(shape = 2, scale = 2)$, $X_3 \sim U(0,5)$, $X_4\sim Gamma(shape = 3, scale = 1)$, $X_5\sim U(0,2)$, $X_6\sim Gamma(shape=1, scale =1)$. Treatment assignment is not randomized and instead depends on the baseline covariates. Specifically, we let $p_G = logit^{-1}(0.2 X_1 + 0.3 X_2 + 0.5 X_3 + 0.2X_4 + 0.4X_5 + 0.1 X_6)$, and then $G\sim Bernoulli(0.5 p_G)$. Note that this construction allows treatment assignment to depend on $\bX$, but in such a way that we can easily tune the overall proportion assigned to treatment versus control, via the coefficient that is multiplied by $p_G$. In our simulation settings, this resulted in roughly equally sized treatment and control groups.

In Setting 1, the surrogate is given by $S = 1.5G + 0.2X_1 + 0.2X_2 + 0.3X_3 + 0.1X_4 + 0.4X_5+0.3X_6 + N(0, 0.4+1.4G)$. In Settings 2-4, the surrogate is given by $S = 1G + 0.2X_1 + 0.2X_2 + 0.3X_3 + 0.1X_4 + 0.4X_5+0.3X_6 + N(0, 0.4+1.4G)$. In Setting 1, the linear model  holds and thus, should perform well. Specifically, the primary outcome is given by $Y = G + 2S + 0.2X_1 + 0.5X_2 + 0.2X_3 + 0.1X_4 + 0.3X_5 + 0.4 X_6 + 2 G X_1 + N(0,1)$. In Setting 2, the linear model is no longer correct, but the terms are additive and therefore the GAM should perform well. The primary outcome for Setting 2 is given by $Y = G + 2S + sin(X_1) + cos(X_2) + X_3^2 + X_4 + log(X_5 + 1) + \sqrt{X_6} + 1.5 G X_1^2 + N(0,1)$. In Setting 3, the terms are no longer additive and thus neither the linear nor the GAM assumptions hold. Specifically, the primary outcome for Setting 3 is given by $Y = G + 2S + 0.5 X_1 X_5^2 + log(X_2 / X_3) + 2 sin(X_4+X_6) + 1.5 G X_1^2 + N(0,1)$. The primary outcome $Y$ in Setting 4 is the same as Setting 1, but without the interaction term for $G$ and $X_1$, so there is no heterogeneity in the PTE. That is, in Setting 4, $Y = G + 2S + 0.2X_1 + 0.5X_2 + 0.2X_3 + 0.1X_4 + 0.3X_5 + 0.4 X_6 + N(0,1)$. The results from Setting 4 are included below in Figure \ref{setting4_plot} and Table \ref{setting4_res}. Similar to the results in the main paper, we see strong performance in terms of coverage levels close to 95\% and small MSE for all choices of base learners, and higher standard errors for the regression forests compared to the linear and GAM base learners.

\clearpage
\begin{figure}
    \centering 
    \includegraphics[scale = 0.08]{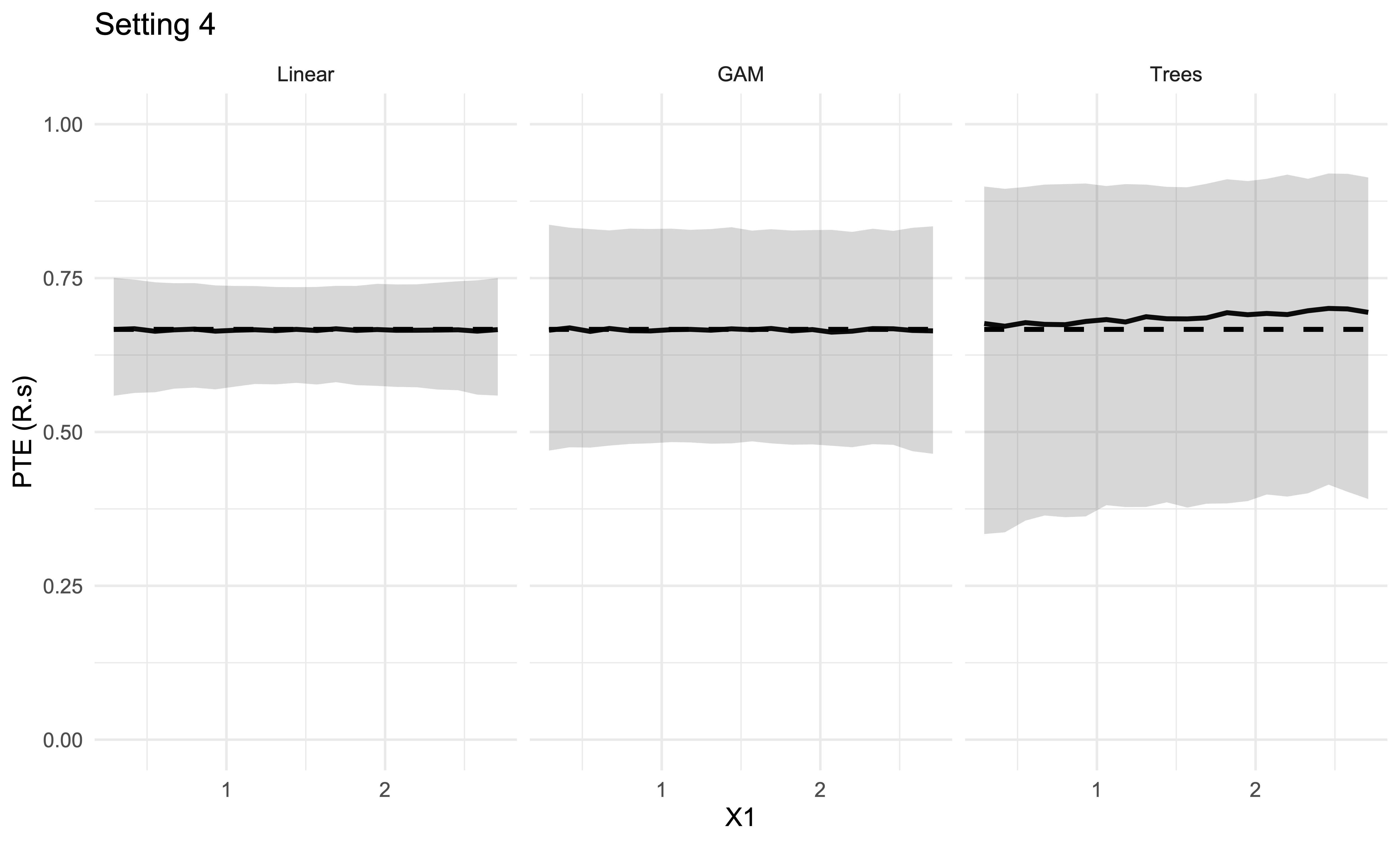}
    \caption{
    Estimated $R_S(\bx)$ (solid lines) vs. true $R_S(\bx)$ (dashed lines) plotted against $X_1$ for Setting 4, which features no heterogeneity in the PTE. Confidence bands (grey shading) obtained using bootstrapping. \label{setting4_plot}
    } 
\end{figure}

\clearpage
\begin{table}[hptb]
\caption{Simulation results for $R_S(\bx)$ in Setting 4, averaged over $X_1$, where Bias reflects the absolute value of the difference between the estimate and the truth, summarized as the median of 1000 iterations; ESE represents the empirical standard error (calculated as the median absolute deviation of estimates across iterations); ASE represents the average standard error (calculated as the median of the bootstrap variance estimates);  MSE represents the median squared error; Coverage indicates the coverage rate of 95\% bootstrap confidence intervals with respect to the truth. \label{setting4_res}}
\begin{center}
\begin{tabular}{|l|c|c|c|c|c|} \hline
&\multicolumn{3}{c|}{Setting 4}\\ \hline
\multicolumn{1}{|l|}{}&\multicolumn{1}{c|}{Linear}&\multicolumn{1}{c|}{GAM}&\multicolumn{1}{c|}{Trees}\\ \hline 
Bias&0.025&0.036&0.073\\ 
ESE&0.037&0.054&0.106\\ 
ASE&0.043&0.087&0.126\\ 
MSE&0.001&0.001&0.005\\ 
Coverage&0.967&0.988&0.984\\ 
\hline
\end{tabular}
\vspace{3mm}
\end{center}
\end{table}

\clearpage
\section*{Appendix B} 
\begin{landscape}
\begin{figure}
    \centering     
    \includegraphics[scale = 0.3]{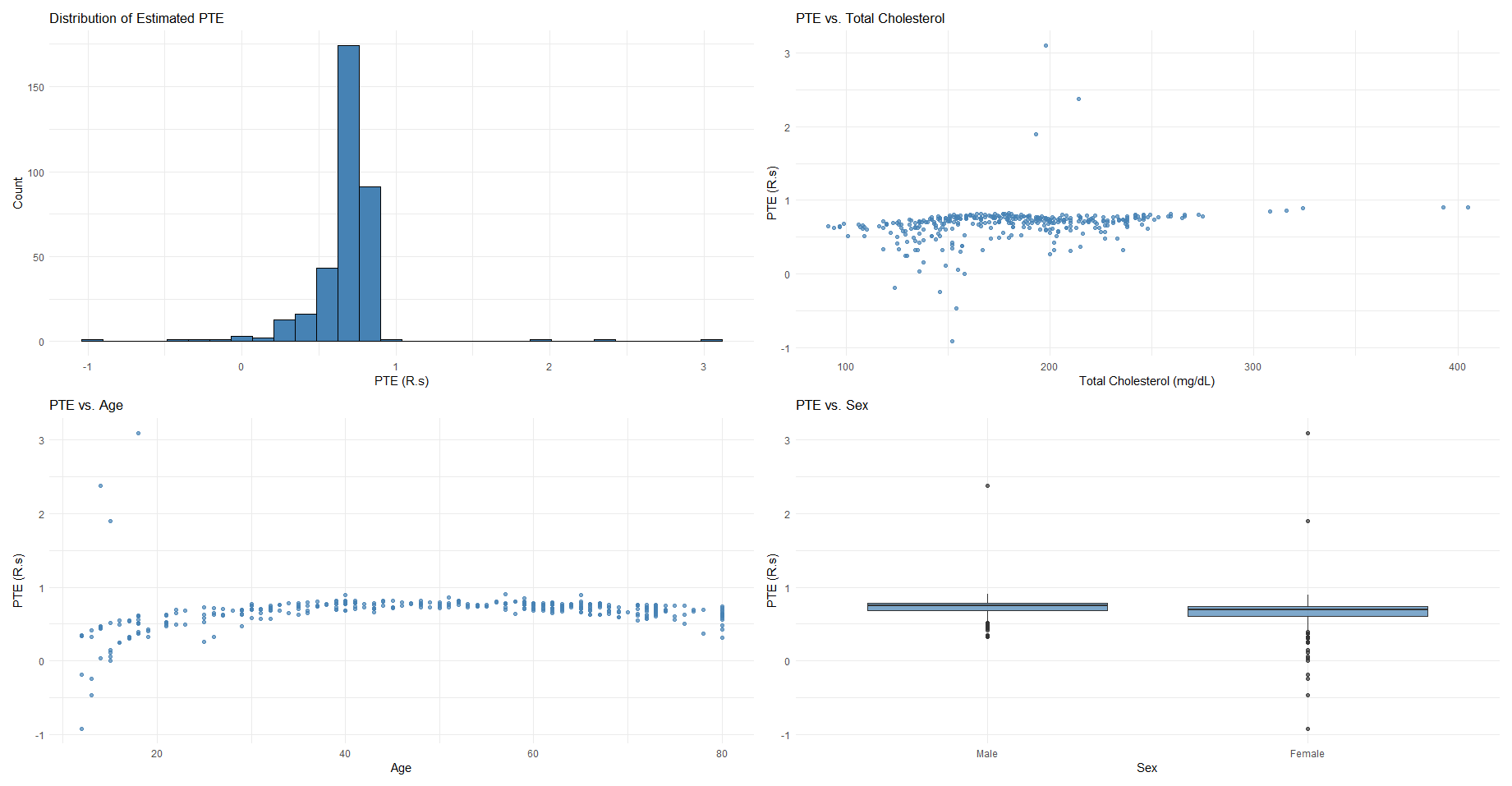}

    \caption{
    Estimation results for the NHANES survey data using GAMs as the base learners, evaluating the strength of HbA1c as a surrogate marker for plasma fasting glucose, when the exposure is obesity status; subfigures show the distribution of PTE estimates (top left panel) and PTE estimates by cholesterol (top right), age (bottom left), and sex (bottom right).
    } \label{nhanes_plot_gam}
\end{figure}
\end{landscape}

\begin{landscape}
\begin{figure}
    \centering     
    \includegraphics[scale = 0.3]{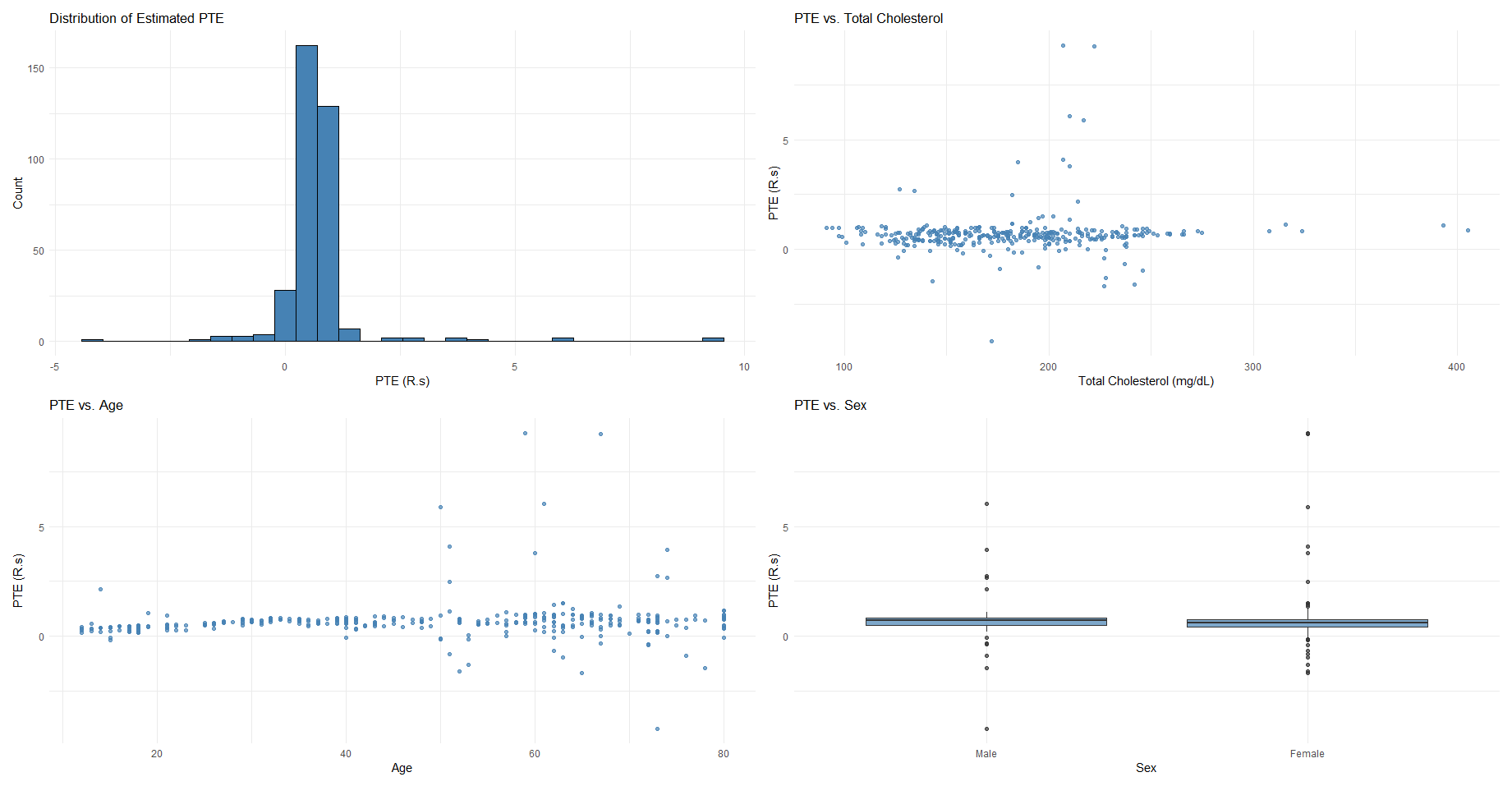}

    \caption{
    Estimation results for the NHANES survey data using regression forests as the base learners, evaluating the strength of HbA1c as a surrogate marker for plasma fasting glucose, when the exposure is obesity status; subfigures show the distribution of PTE estimates (top left panel) and PTE estimates by cholesterol (top right), age (bottom left), and sex (bottom right).
    } \label{nhanes_plot_trees}
\end{figure}
\end{landscape}

Our main example in Section \ref{example} uses linear base learners to illustrate our proposed approach; here we examine using GAMs and regression forests applied to the same NHANES data set, with results shown in Figures \ref{nhanes_plot_gam} and \ref{nhanes_plot_trees}, respectively. The estimated PTE remained high across methods, suggesting that the surrogate is strong for most patients. However, GAMs and regression forests, particularly the latter, produced more outliers in the PTE estimates (the regression forest results contained one extreme outlier that was removed for better visualization) compared to the results using linear base learners, consistent with our simulation findings where linear base learners produced fewer extreme estimates due to their more constrained form. 

The suggested relationship between PTE and total cholesterol was similar across the different learners, with PTE increasing slightly as cholesterol increases. The relationship between PTE and sex was less pronounced in GAMs and regression forests compared to linear models. The most notable difference appeared in the age covariate: while linear base learners showed higher PTE at younger ages, regression forests showed the opposite trend, and GAMs indicated higher PTE for middle-aged subjects with lower values for both young and old individuals.

Without knowing the ground truth, we cannot determine which choice of learners best captures reality. Based on our simulations, linear base learners may miss complex patterns in specific covariate regions due to their rigid structure, while regression forests offer flexibility but potentially less stability. GAMs' strong performance in our simulations suggests they may be capturing a more nuanced age relationship that linear models missed. In practice, clinical expertise should guide base learner selection based on expected relationships.

\clearpage 

\clearpage 

\end{document}